\documentclass[pdflatex,sn-mathphys]{sn-jnl}
\usepackage{enumerate}

\jyear{2022}%

\theoremstyle{thmstyleone}%
\newtheorem{theorem}{Theorem}
\theoremstyle{thmstyletwo}%

\theoremstyle{thmstylethree}%
\newtheorem{problem}{Problem}%

\begin{document}

\title[Quantum All-Subkeys-Recovery Attacks on 6-round Feistel-2*]{Quantum All-Subkeys-Recovery Attacks on 6-round Feistel-2* Structure Based on Multi-Equations Quantum Claw Finding}

\author*[1,2]{\fnm{Wenjie} \sur{Liu}}\email{wenjiel@163.com}

\author[2]{\fnm{Mengting} \sur{Wang}}\email{wmt$\_$de$\_$wyyx@163.com}

\author[2]{\fnm{Zixian} \sur{Li}}\email{zixianli157@163.com}

\affil[1]{\orgdiv{Engineering Research Center of Digital Forensics, Ministry of Education}, \orgname{Nanjing University of Information Science  and Technology},
\orgaddress{\city{Nanjing}, \postcode{210044}, \state{Jiangsu} \country{China}}}
\affil[2]{\orgdiv{School of Computer Science}, \orgname{Nanjing University of Information Science  and Technology},
\orgaddress{\city{Nanjing}, \postcode{210044}, \state{Jiangsu} \country{China}}}


\abstract{Exploiting quantum mechanisms, quantum attacks have the potential ability to break the cipher structure. Recently, Ito et al. proposed a quantum attack on Feistel-2* structure (Ito et al.'s attack) based on the Q2 model. However, it is not realistic since the quantum oracle needs to be accessed by the adversary, and the data complexity is high. To solve this problem, a quantum all-subkeys-recovery (ASR) attack based on multi-equations quantum claw-finding is proposed, which takes a more realistic model, the Q1 model, as the scenario, and only requires 3 plain-ciphertext pairs to quickly crack the 6-round Feistel-2* structure. First, we proposed a multi-equations quantum claw-finding algorithm to solve the claw problem of finding multiple equations. In addition, Grover's algorithm is used to speedup the rest subkeys recovery. Compared with Ito et al.'s attack, the data complexity of our attack is reduced from $O(2^n)$ to $O(1)$, while the time complexity and memory complexity are also significantly reduced.}

\keywords{Quantum attack, All-subkeys-recovery attack, 6-round Feistel-2*, Multi-equations quantum claw finding, Grover's algorithm}



\maketitle

\section{Introduction}\label{sec1}

With the rapid development of quantum computing, classical cryptography has been gradually threatened. In 1994, Shor's algorithm \cite{Shor1999} was proposed to solve the big integer factorization problem, which makes the current public key systems, such as RSA, ECC, face great challenges. In 1995, Grover \cite{Grover1996} proposed a quantum fast search algorithm, namely the well-known Grover's algorithm. It turns a problem that requires complexity $O(N)$ on a classical computer into one that can be completed in complexity $O(\sqrt N )$ time, which greatly reduces the computational complexity of the problem. Subsequently, some scholars have improved Grover's algorithm \cite{Long2001} \cite{Toyama2013}, and even applied them  to crack symmetric cryptosystems \cite{Gregor2017}.

In the cryptanalysis of symmetric ciphers, the introduction of quantum computing has become a hot research topic in recent years. In 2010, Kuwakado et al. proposed a quantum 3-round distinguisher of Feistel construction \cite{Hidenori2010} based on the Simon's algorithm \cite{Daniel1997}, which reduced the time complexity of key-recovery from O($2^n$) to O($n$). This also enables more and more concrete structures to be evaluated, such as Even-Mansour cipher \cite{Hidenori2012}, CBC-like MACs \cite{Kaplan2016a, Santoli2017}, AEZ \cite{Shi2018}, AES-COPA \cite{Xu2021}, FX construction \cite{Gregor2017}, Feistel constructions \cite{Xavier2020, Dong2018, Dong2019, Dong2020, Hosoyamada2018} et al. Kaplan \cite{Kaplan2016b} divides these quantum attacks into two attack models according to the opponent's ability, i.e. the Q1 model and the Q2 model.\\
\textbf{Q1 model}: The adversary query the oracle through classical methods and the processing of the data can be done using quantum computers.\\
\textbf{Q2 model}: The adversary query the oracle through online quantum superposition and the processing of the data can be done using quantum computers. \\
It can be seen that the adversary in Q1 model is much more realistic than those in Q2 model. This means that if the adversary is not permitted to access the quantum cipher oracle, the attack in the Q2 model will not work.

Among these quantum attacks, the quantum key-recovery attack on the r-round Feistel structure is a hot topic. In block cryptography, the design method of the Feistel construction \cite{Feistel1973} is widely used, such as DES \cite{Coppersmith1994}, Camellia \cite{Aoki2001}, and CAST \cite{Adams1997}, all use the Feistel construction. In 2017, Yang et al. \cite{Yang2017} applied differential function reduction technique on the 6-round Feistel-2* ciphers in classical attack, which reduces the time complexity to $O(5\cdot{2^{n/2+1}})$. In 2019, Ito et al. \cite{Ito2019} introduced the first 4-round quantum distinguisher on Feistel ciphers in the quantum chosen-ciphertext setting (Q2 model) and extended the distinguisher to Feistel-2* structures for key recovery attacks. However, this attack belongs to the Q2 model might appear rather extreme and perhaps even unrealistic since the adversary needs to be allowed access to the quantum oracle. In addition, it requires a large amount of plain-ciphertext pairs, which has a high data complexity.

In this paper, a quantum ASR attack based on multi-equations quantum claw-finding is proposed, which can recover all subkeys of the 6-round Feistel-2* structure with only 3 plain-ciphertext pairs. First, a multi-equations quantum claw finding algorithm is proposed to find a claw $(x_1, x_2)$ between the functions $f_i$ and $g_i$ to make $f_i(x_1) = g_i(x_2)$. In addition, Grover's algorithm is used to speedup the rest subkeys recovery. With such a method, a quadratic accelerated search can be performed compared to an exhaustive algorithm. The complexity analysis shows that the time complexity of our attack is $O(2^{n/3})$, the data complexity is $O(1)$ and the number of qubits is $O(n)$. Compared with the other attack, our attack not only has lower time complexity but also has lower data complexity.

The rest of this paper is organized as follows. Sect. \ref{sec2} covers some basics including notations, Feistel-2* structure, quantum claw finding algorithm and Grover's algorithm. In Sect. \ref{sec3}, a multi-equations quantum claw search algorithm is proposed. In Sect. \ref{sec4}, we review the quantum chosen-plaintext attacks against Feistel-2*  and  use our proposed algorithm to attack the 6-round Feistel-2* structure. Then, a complexity analysis with classical and the existing quantum attacks is provided in Sect. \ref{sec5}. Finally, the conclusion is given in Sect. \ref{sec6}.

\section{Preliminaries}\label{sec2}

\subsection{Notations}\label{subsec2.1}

Some notations and their descriptions that will be used in this paper are shown in Tab. \ref{tab1}.
\begin{table}[h]
\begin{center}
\caption{Some notations and their descriptions}\label{tab1}%
\begin{tabular}{@{}ll@{}}
\toprule
Notation & Description \\
\midrule
$n$    & Block size    \\
$L_i, R_i$    & The left and right inputs of the $i$-th round    \\
$F_i$        & The $i$-th round function (a mapping from $\left\{{0,1}\right\}^{n/2}$ to $\left\{{0,1}\right\}^{n/2}$)     \\
$K_i$        & The subkey of the $i$-th round ($n/2$ bit)     \\
$a\lvert b$       & Connection     \\
\toprule
\end{tabular}
\end{center}
\end{table}

\subsection{Feistel-2* structure}\label{subsec2.2}

\begin{figure}
    \centering
    \includegraphics[width=9cm]{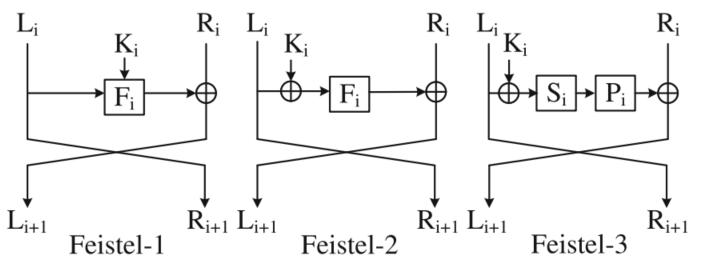}
    \caption{Three Feistel constructions}
    \label{fig1}
\end{figure}

The Feistel structures are divided into three categories \cite{Isobe2013} as shown in Fig. \ref{fig1}. Feistel-1 is a Feistel cipher with a random key value F function and assumes that its subkeys are independent. The round function of Feistel-2 is composed of a common F function and a subkey XORed before the F function. Feistel-3 is a subset of Feistel-2, and its F-function consists of a bijective S-box layer and a linearly permuted P layer.

\begin{figure}
    \centering
    \includegraphics[width=4cm]{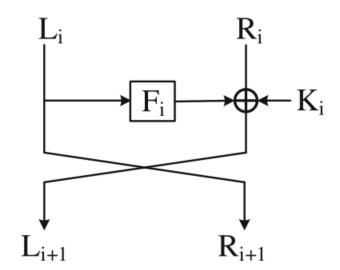}
    \caption{Feistel-2* construction}
    \label{fig2}
\end{figure}

Feistel-2* is a variant of the Feistel-2 cipher, which is shown in Fig. \ref{fig2}. The round function of Feistel-2 consists of a common F function and a subkey XORed before the F function, while the subkey of Feistel-2* is XORed after the F function. Feistel-2* is widely used in proposals, such as SIMON and Simeck. Specifically, if $L_i$ and $R_i$ are the left input and right input of the $i$-th round of Feistel-2*, respectively, the calculations of the left output $L_{i+1}$ and the right output $R_{i+1}$ of the $i$-th round of Feistel-2* are as follows:
\begin{equation}
    L_{i+1}=R_i+F_i(L_i)+K_i, \quad\quad\quad    
    R_{i+1}=L_i .
\end{equation}

\subsection{Quantum claw finding algorithm}\label{subsec2.3}

We introduce the claw finding problem.

\begin{problem}[\textbf{Claw finding}]\label{pro2}
Suppose there are two functions $f:{\left\{ {0,1} \right\}^u} \to {\left\{ {0,1} \right\}^v}$ and $g:{\left\{ {0,1} \right\}^u} \to {\left\{ {0,1} \right\}^v}$, if there exists a pair $\left( {x_1,x_2} \right) \in {\left\{ {0,1} \right\}^u} \times {\left\{ {0,1} \right\}^u}$ such that $f\left( x_1 \right){\rm{ = }}g\left( x_2 \right)$, this pair is called a \textbf{claw} of the functions $f$ and $g$. the problem is to find the claw.
\end{problem}

Ambainis \cite{Ambainis2017} proposed a quantum walk algorithm (as shown in Algorithm \ref{algo1}). In Algorithm \ref{algo1}, the state $\lvert {S,{x_S}} \rangle$ is the basis state we use, where $S=\left\{i_1,i_2,\cdots,i_r\right\}$ is a subset of $r$ indexes, and $x_S=\left\{x_{i_1},x_{i_2},\cdots,x_{i_r}\right\}$ is the values corresponding to $S$. Each transformation in Step 1 or 3 is a variant of the ``diffusion transformation" \cite{Grover1996}, and each query in Step 2 or 4 is an oracle $O$ that maps $\lvert {i,a,z} \rangle$ to $\lvert {i,(a+x_i)}  mod  M, z \rangle$, where $i$ consists of $\left\lceil {logN} \right\rceil $  bits, a consists of $\left\lceil {logM} \right\rceil $ quantum bits and $z$ consists of all other bits. Each subset $S$ is regarded as a vertex, and by a quantum walk we move from one vertex $S$ to another vertex $S'$. By several walks, we will get closer and closer to a subset marked before.

On the basis of Ambainis, Zhang \cite{Zhang2005} proposed a quantum claw finding algorithm (as shown in Algorithm \ref{algo2}). The idea is to perform quantum walks on two subsets $S_1\subseteq J_1$ and $S_2\subseteq J_2$ in two subsets $J_1\subseteq [N]$ and $J_2\subseteq [N]$ respectively. In Algorithm~\ref{algo2}, Step 3 is to perform a phase-flip unitary operator, which marks all subset $S_1,S_2$ where the claw $(j_1, j_2)\in S_1\times S_2$. Zhang's algorithm satisfies Theorem \ref{thm2}.

\begin{algorithm}
\caption{Quantum Walk on S in A \cite{Andris2007}}\label{algo1}
\begin{algorithmic}[1]
\Require  State $\lvert {S,{x_S},i} \rangle $ and $A$ with $S \subseteq A$, and $i \in A - S$. Let $\lvert S \lvert = r$, $\lvert A \lvert = N$, where ${{x_S}}$ contains all variable values ${{x_j}}$ of $j \in S$
\State $\lvert {S,{x_S},i} \rangle  \to \lvert {S,{x_S}} \rangle \left( {\left( {{\rm{ - }}1{\rm{ + }}\frac{2}{{N - r}}} \right)\lvert i \rangle  + \frac{2}{{N - r}}\sum\nolimits_{j \in A - S - \{ i\} } {\lvert j \rangle } } \right)$
\State $\lvert {S,{x_S},i} \rangle  \to \lvert {S \cup \{ i\} ,{x_{S \cup \{ i\} }},i} \rangle $ by one query.
\State $\lvert {S,{x_S},i} \rangle  \to \lvert {S,{x_S}} \rangle \left( {\left( {{\rm{ - }}1{\rm{ + }}\frac{2}{{r+1}}} \right)\lvert i \rangle  + \frac{2}{{r+1}}\sum\nolimits_{j \in S - \{ i\} } {\lvert j \rangle } } \right)$
\State $\lvert {S,{x_S},i} \rangle  \to \lvert {S - \{ i\} ,{x_{S - \{ i\} }},i} \rangle $ by one query.
\end{algorithmic}
\end{algorithm}

\begin{algorithm}
\caption{Quantum claw finding algorithm \cite{Zhang2005}}\label{algo2}
\begin{algorithmic}[1]
\Require  ${x_1},...,{x_N} \in [M]$. ${J_1},{J_2} \subseteq [N]$, $\lvert  {{J_1}} \lvert  = m$, $\lvert  {{J_2}} \lvert  = n$. $R$ is the Equality relation  s.t. there is at most one $\left( {{x_{{j_1}}},{x_{{j_2}}}} \right) \in R$ with ${j_1} \subseteq {J_1}$, ${j_2} \subseteq {J_2}$ and ${j_1} \ne {j_2}$.
\Ensure $\left( {{j_1},{j_2}} \right)$ if they exist; otherwise reject.
\State {Set the initial state be 

\noindent$\lvert {{\psi _{start}}} \rangle  = \frac{1}{{\sqrt T }}\sum\nolimits_{{S_b} \subseteq {J_b},\lvert  {{S_b}} \lvert  = {r_b},{i_b} \in {J_b} - {S_b}} {\lvert{{S_1},{x_{{S_1}}},{i_1},{S_2},{x_{{S_2}}},{i_2}}\rangle }$, where $T = \left( \begin{array}{l}
m\\
{r_1}
\end{array} \right)\left( \begin{array}{l}
n\\
{r_2}
\end{array} \right)\left( {m - {r_1}} \right)\left( {n - {r_2}} \right)$ and $b = 1,2$
.}
\State Repeat $\Theta \left( {\sqrt {\frac{{mn}}{{{r_1}{r_2}}}} } \right)$ times:
\State \quad \quad {Do the phase flip $\lvert  {{S_1},{x_{{S_1}}},{i_1},{S_2},{x_{{S_2}}},{i_2}} \rangle  \to  - \lvert {{S_1},{x_{{S_1}}},{i_1},{S_2},{x_{{S_2}}},{i_2}}\rangle $ 

while the unique $\left( {{j_1},{j_2}} \right)$ is in ${S_1} \times {S_2}$.}
\State \quad \quad {Do \textbf{Quantum Walk} on ${S_1}$ in ${{J_1}}$ for ${t_1} = \left\lceil {\frac{\pi }{4}\sqrt {{r_1}} } \right\rceil $ times.

\State \quad \quad Do \textbf{Quantum Walk} on ${S_2}$ in ${{J_2}}$ for ${t_2} = \left\lceil {\frac{\pi }{4}\sqrt {{r_2}} } \right\rceil $ times.}

\State Measure the resulting state and give the corresponding answer.
\end{algorithmic}
\end{algorithm}

\begin{theorem}[]\label{thm2}
Algorithm \ref{algo2} outputs desired results correctly in the function evaluation, and we can pick $r_1$ and $r_2$ to make number of queries be:
\[\left\{ \begin{array}{l}
O\left( {{{\left( {mn} \right)}^{1/3}}} \right)\begin{array}{*{20}{c}}
{}&{if}
\end{array}\sqrt n  \le m \le {n^2}\begin{array}{*{20}{c}}
{}&{}
\end{array}\left(by\ letting\quad {{r_1} = {r_2} = {{\left( {mn} \right)}^{1/3}}} \right)\\
\\
O\left( {\sqrt n } \right)\begin{array}{*{20}{c}}
{}&{}&{}&{}&{}&{}&{}&{if}
\end{array}m \le \sqrt n \begin{array}{*{20}{c}}
{}&{}{}&{}&{}{}&{}&{}{}&{}&{}
\end{array}\left(by\ letting\quad  {{r_1} = m,{r_2} \in \left[ {m,{{\left( {mn} \right)}^{1/3}}} \right]} \right)\\
\\
O\left( {\sqrt m } \right)\begin{array}{*{20}{c}}
{}&{}&{}&{}&{}&{}&{if}
\end{array}m > {n^2}\begin{array}{*{20}{c}}
{}&{}{}&{}&{}{}&{}&{}{}&{}&{}&{}
\end{array}\left(by\ letting\quad  {{r_1} \in \left[ {n,{{\left( {mn} \right)}^{1/3}}} \right],{r_2} = n} \right)
\end{array} \right.\]
\end{theorem}

\subsection{Grover's algorithm}\label{subsec2.4}

Given a set $N$, in which some elements are marked, the goal is to find a marked element from $N$. We use $M \subseteq N$ to denote the marker element. Classically, it takes a time of $\lvert N \lvert/\lvert M \lvert$ to solve this problem. However, in a quantum computer, this problem has a high probility to be solved in a time of $\sqrt {\lvert N \lvert/\lvert M \lvert} $ by using Grover's algorithm. The steps of the algorithm are as follows:

(1) Initialization of an $n$-bit register ${\lvert 0 \rangle ^{ \otimes n}}$ and applying the Hadamard transform to the first register.
\begin{equation}
    {H^{ \otimes n}}\lvert 0\rangle  = \frac{1}{{\sqrt {{2^n}} }}\sum\limits_{x \in {{\left\{ {0,1} \right\}}^n}} {\lvert x \rangle }  =\lvert \varphi  \rangle  
\end{equation}

(2) Construct an oracle ${O_B}:\lvert x \rangle  \to {\left( { - 1} \right)^{B\left( x \right)}}\lvert x \rangle $, where $B\left( x \right) = 1$ if $x$ is the correct solution; otherwise, $B\left( x \right) = 0$.

(3) Apply Grover‘s iteration $R \approx \frac{\pi }{4}\sqrt {\lvert N \lvert/\lvert M \lvert} $ times that can be given as follows:
\begin{equation}
    {\left[ {\left( {2\lvert \varphi  \rangle \langle \varphi  \lvert - I} \right){O_B}} \right]^{R}}\lvert \varphi  \rangle  \approx \lvert {{x_0}} \rangle 
\end{equation}

(4) Obtain $x_0$ by measuring the result state.

Theoretically, when the number of iterations is around $\frac{\pi }{4}\sqrt {\lvert N \lvert/\lvert M \lvert}$, the probability of finding the target item in Grover's algorithm \cite{Grover1998} is optimal.

\section{Multi-equations quantum claw finding algorithm}\label{sec3}

If there are two functions $f:{\left\{ {0,1} \right\}^u} \to {\left\{ {0,1} \right\}^v}$ and $g:{\left\{ {0,1} \right\}^u} \to {\left\{ {0,1} \right\}^v}$, and a claw $\left( {x_1,x_2} \right) \in {\left\{ {0,1} \right\}^u} \times {\left\{ {0,1} \right\}^u}$ satisfying $f\left( x_1 \right) = g\left( x_2 \right)$, we can use Quantum claw finding algorithm (see Algorithm \ref{algo2}) to find the claw $\left( {x_1,x_2} \right)$. However, if there are $w$ functions $f_{i}:{\left\{ {0,1} \right\}^u} \to {\left\{ {0,1} \right\}^v}$ and $g_{i}:{\left\{ {0,1} \right\}^u} \to {\left\{ {0,1} \right\}^v}$, $1\le i \le w$, and a claw $\left( {x_1,x_2} \right) \in {\left\{ {0,1} \right\}^u} \times {\left\{ {0,1} \right\}^u}$ satisfying $\forall 0\le i\le w, f_{i}\left( x_1 \right) = g_{i}\left( x_2 \right)$, then Algorithm \ref{algo2} doesn't work anymore. To solve this problem, a multi-equations quantum claw finding algorithm is proposed.

We connect ${f_1}\left( x_1 \right)$, ${f_2}\left( x_1 \right), ..., {f_{i}}\left( x_1 \right)$ in series into a function $f\left( x_1 \right) = {f_1}\left( x_1 \right)\lvert \lvert {f_2}\left( x_1 \right)\lvert \lvert... \lvert \lvert {f_{i}}\left( x_1 \right)$, and ${g_1}\left( x_2 \right)$, ${g_2}\left( x_2 \right) ..., {g_{i}}\left( x_2 \right)$ into a function $g\left( x_2 \right) = {g_1}\left( x_2 \right)\lvert \lvert {g_2}\left( x_2 \right)\lvert \lvert...\lvert \lvert {g_{i}}\left( x_2 \right)$. In fact, Algorithm~\ref{algo2} cannot directly solve the multi equations claw finding problems after the functions are connected in series, because the conditions of its function domain do not meet the requirements of ${J_1} \cap {J_2} = \emptyset $. Therefore, we construct the function $F:{\left\{ {0,1} \right\}^{u+1}} \to {\left\{ {0,1} \right\}^{vw}}$, 
\begin{equation}
F(c\lvert\lvert x)  = \left\{ \begin{array}{l}
f(x),\begin{array}{*{20}{c}}
{}&{}
\end{array}if\begin{array}{*{20}{c}}
{}&{}
\end{array}c= 0\\
g(x),\begin{array}{*{20}{c}}
{}&{}
\end{array}if\begin{array}{*{20}{c}}
{}&{}
\end{array}c= 1\\
\end{array} \right.
,\end{equation}
where the additional qubit $c$ controls the selection of functions $f$ and $g$. At this time, their function domains are $J_1 = \{0,1,\cdots,N-1 \}$ and $J_2 = \{N,N+1,\cdots,2N-1 \}$ respectively, that is ${J_1} \cap {J_2} = \emptyset $.

To implement the algorithm, we define a quantum query as an oracle $Q_F$ that maps $\lvert {x,y,z} \rangle$ to $\lvert {x,(y+F(x))}  \mod {M}, z \rangle$. Set $N=2^{u}$, $M=2^{vw}$, $m=n=2^{u}$, $J_1 = \{0,1,\cdots,N-1 \}$, $J_2 = \{N,N+1,\cdots,2N-1 \}$, $r_1=r_2={\left(mn\right)}^{\frac{1}{3}}=2^{\frac{2u}{3}}$, and the relation $R$ to be $F\left( {{j_1}} \right) = F\left( {{j_2}} \right)$, then perform the quantum claw finding. Consequently, a multi-equation quantum claw finding algorithm is proposed, and the specific procedure is given in Algorithm~\ref{algo3}. In addition, the above ``diffusion transform'' is used, as $D_1$ and $D_2$ in Algorithm~\ref{algo3}.

\begin{algorithm}
\caption{Multi-equation quantum claw finding algorithm }\label{algo3}
\begin{algorithmic}[1]
\Require  $f_{i}:{\left\{ {0,1} \right\}^u} \to {\left\{ {0,1} \right\}^v}$ and $g_{i}:{\left\{ {0,1} \right\}^u} \to {\left\{ {0,1} \right\}^v}$, $1\le i \le w$.

\Ensure Claw $\left( {{x_1},{x_2}} \right)$, or reject if it does not exist. 

\State Connect functions $f_{i}$, $g_{i}$, respectively, and get $f\left( x \right) = {f_1}\left( x \right)\lvert \lvert {f_2}\left( x \right)\lvert \lvert... \lvert \lvert {f_{i}}\left( x \right)$ and $g\left( x \right) = {g_1}\left( x \right)\lvert \lvert {g_2}\left( x \right)\lvert \lvert  ...$ $\lvert \lvert {g_{i}}\left( x \right)$, $1\le i \le w$. 

\State Construct $F:{\left\{ {0,1} \right\}^{u+1}} \to {\left\{ {0,1} \right\}^{vw}}$ as
\begin{equation} \nonumber
F(c\lvert\lvert x)  = \left\{ \begin{array}{l}
f(x),\begin{array}{*{20}{c}}
{}&{}
\end{array}if\begin{array}{*{20}{c}}
{}&{}
\end{array}c= 0\\
g(x),\begin{array}{*{20}{c}}
{}&{}
\end{array}if\begin{array}{*{20}{c}}
{}&{}
\end{array}c= 1\\
\end{array} \right.
,\end{equation}
where $c$ controls the selection of
functions $f$ and $g$.



\State {Prepare the initial state

\noindent \begin{equation}  \nonumber
\frac{1}{{\sqrt T }}\sum\nolimits_{{S_b} \subseteq {J_b},\lvert  {{S_b}} \lvert  = {r},{z_b} \in {J_b} - {S_b}, b\in\{1,2\}} {\lvert S_1\rangle_{s_1}\lvert z_1 \rangle_{t_1}  \lvert S_2 \rangle_{s_2}\lvert z_2 \rangle_{t_2} }\lvert 0 \rangle_{f_1}\lvert 0\rangle_{f_2}\lvert 0 \rangle_{a_1} \lvert 0 \rangle_{a_2} , \end{equation}

\noindent where $T = \left(C_{N}^{r}\left( {N - r} \right)\right)^2$, $N = 2^{u}$, $r = N^{\frac{2}{3}}$, and $J_1 = \{0,1,\cdots,N-1 \}$, $J_2 = \{N,N+1,\cdots,2N-1 \}$. For convenience, we divide the register of state into $s_b$, $f_b$, $t_b$, $a_b$ $(b=1,2)$, which hold $r(u+1),rvw,u+1,u+1+vw$ qubits respectively.} 

\State {Perform a unitary operator ${Q_F}:\left\lvert j \right\rangle \lvert y \rangle  \to \left\lvert j \right\rangle \left\lvert {y\oplus F(j)} \right\rangle $ ($j\doteq c\lvert \lvert x$) on registers $(s_b,f_b)$ respectively:

\noindent \begin{equation} \nonumber \lvert S_b\rangle_{s_b} \lvert0 \rangle_{f_b}\to\lvert S_b,F(S_b) \rangle_{s_{b}f_b}. \end{equation}}

\For{$k=0$ \textbf{to} $\Theta \left( \frac{N}{r} \right)$} 
    \If{$F\left( {{j_1}} \right) = F\left( {{j_2}} \right)$ and $ {{j_1}\doteq(0\lvert \lvert x_1)} \in {S_1}$, $ {{j_2}\doteq(1\lvert \lvert x_2)} \in {S_2}$}
        \State {Flips the phase of the state on $h_1, h_2$ 
        
       \noindent $\lvert {S_1}\rangle_{s_1}\lvert{F(S_1)}\rangle_{f_1}\lvert{z_1}\rangle_{t_1}\lvert{S_2}\rangle_{s_2}\lvert{F(S_2)}\rangle_{f_2}\lvert{z_2}\rangle_{t_2}\to$$ -\lvert{S_1},{F(S_1)},{z_1}\rangle_{h_1}\lvert{S_2},{F(S_2)},{z_2}\rangle_{h_2}$,
       
        \quad where $h_1=(s_1,f_1,t_1),h_2=(s_2,f_2,t_2)$.
        }
    \EndIf
  
    \For{$k=0$ \textbf{to} $\left\lceil {\frac{\pi }{4}\sqrt {r} } \right\rceil$}
        \For{$b=1$ \textbf{to} $2$} 
            \State {Perform operator $D_1^{(b)}$ on $h_b$
            
            $\lvert{S_b,{F(S_b)},z_b} \rangle_{h_b}\to$$\lvert {S_b,{F(S_b)}} \rangle_{s_bf_b}\left[{( {{\rm{ - }}1{\rm{+}}\frac{2}{{N - r}}} )\lvert z_b \rangle+ \frac{2}{{N - r}}\sum\nolimits_{z'_b\in J_b- S_b-\{z_b\}} {\lvert z'_b \rangle } } \right]_{t_b}$.}
                    
            \State {Query $F(z_b)$ by using operator $Q_F$
            
            $\lvert {S_b,{F(S_b)},z_b} \rangle_{h_b}  \lvert 0 \rangle_{a_b}   \rightarrow  \lvert {S_b \cup \{ z_b\} ,{F(S_b \cup \{ z_b\}) },z_b} \rangle _{h_{b}a_b}$.}
                     
            \State {Perform operator $D_2^{(b)}$ on $h_b$
            
            $\lvert {S_b,{F(S_b)},z_b} \rangle_{h_{b}a_b}   \rightarrow\lvert {S_b,{F(S_b)}} \rangle_{s_{b}f_{b}a_{b}}\left[ {( {{\rm{ - }}1{\rm{ + }}\frac{2}{{r+1}}})\lvert z_b \rangle  + \frac{2}{{r+1}}\sum\nolimits_{z'_b \in S_b - \{ z_b\} } {\lvert z'_b \rangle } } \right]_{t_b}$.}
                     
            \State {Query $F(z_b)$ by using operator $Q_F$
            
            $\lvert {S_b,{F(S_b)},z_b} \rangle_{h_{b}a_b}  \rightarrow\lvert {S_b - \{ z_b\} ,{F(S_b - \{ z_b\} )},z_b} \rangle_{h_b}\lvert 0 \rangle_{a_b}  $.}
        \EndFor
    \EndFor
\EndFor
\State Obtain $(S_1,S_2)$ by measuring the resulting state of registers $(s_1,s_2)$.
\If{the solution $(j_1,j_2)\in S_1\times S_2$}
        \State Get $\left( {{x_1},{x_2}} \right)$ from $(j_1,j_2)$.
\Else
        \State Rejects.
\EndIf
\end{algorithmic}
\end{algorithm}

The query complexity of Algorithm~\ref{algo3} is $O(N^{\frac{2}{3}})$ by Theorem~\ref{thm2}. The memory complexity and the number of qubits of it are $O(r)=O(N^{\frac{2}{3}})$ and $O(N^{\frac{2}{3}}\log N)$ respectively to store the state $\left \lvert S_1, F(S_1), z_1 , S_2, F(S_2), z_2\right \rangle$.

\section{Quantum All-Subkeys-Recovery Attacks on Feistel-2* Structure}\label{sec4}

\subsection{Review of quantum chosen-plaintext attacks against Feistel-2*}\label{subsec4.1}

Here we will review the idea of key-recovery attack on Feistel-2* structure by Ito et al \cite{Ito2019}. They construct a 5-round distinguisher following Kuwakado and Morii's 3-round distinguisher \cite{Hidenori2010} by computing the output of the first F function and the last F function in the encryption. They combine this 5-round distinguisher with Grover search to attack the 6-round Feistel-2* structure. \\
\textbf{Attack Strategy.} Their attacks use not only encrypted oracle, but also decrypted oracle. Given the quantum encryption and decryption oracle of the 6-round Feistel-2* structure, run the following steps on the quantum circuit.

1. Implement a quantum circuit $E$. Take the intermediate state value after the first round and the subkey of the first round as input and compute the plaintext by decrypting the first round. Perform a quantum query to oracle on the computed plaintext and return the oracle output.

2. Implement a quantum circuit $D$ and calculate the reciprocal of $E$. It takes the ciphertext and subkey of the first round as input, performs a quantum decryption query on the ciphertext to obtain the plaintext, and calculates the intermediate state value after the first round based on the plaintext and subkey of the first round.

3. Guess the subkey of the first round.

4. For each guess, check its correctness by following the steps below.

\quad(a) Use the 5-round distinguisher to $E$ and $D$.

\quad(b) If the recognizer returns ``This is a random permutation", the guess is incorrect. Otherwise, the guess is judged to be correct.

In the quantum chosen-plaintext setting, they can recover the key of the 6-round Feistel-2* structure in time $O({2^{n/4}})$.

\subsection{Quantum all-subkeys-recovery attacks on 6-round Feistel-2* structure based on multi-equitions quantum claw finding}\label{subsec4.2}

\begin{figure}
    \centering
    \includegraphics[width=8cm]{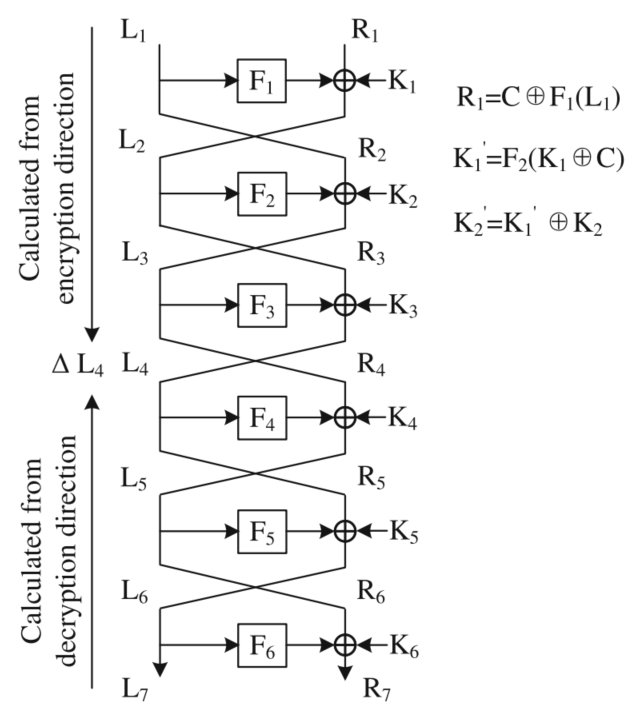}
    \caption{6-round Feistel-2* structure}
    \label{fig3}
\end{figure}

In this subsection,  the quantum ASR attacks on 6-round Feistel-2* structure is proposed, and the structure is shown in Figure \ref{fig3}.

According to classical ASR attack \cite{Yang2017}, assuming that the selected plaintexts are $L_1\lvert R_1$ and $L_1^{'} \lvert R_1^{'}$, the corresponding outputs are $L_{r+1}\lvert R_{r+1}$ and $L_{r+1}^{'} \lvert R_{r+1}^{'}$, where $r \ge 3$. The rule for selecting plaintext is
\begin{equation}
{F_1}\left( {{L_1}} \right) \oplus {R_1} = {F_1}\left( {L_1^{'}} \right) \oplus R_1^{'} = C,
\end{equation}
where $C$ is a constant. Supposing $K_1^{'}$ and $K_2^{'}$ are new subkeys, so its rule is as follows
\begin{equation}
K_1^{'} = {F_2}\left( {{K_1} \oplus C} \right),
\end{equation}
\begin{equation}
K_2^{'} = K_1^{'} \oplus {K_2}
..\end{equation}

According to the calculation rules of the Feistel-2* construction, $L_3$ and $R_3$ are represented as ${L_1} \oplus K_2^{'}$ and $C \oplus {K_1}$, respectively. Then, $L_4$ can be calculated by using $L_3$ and $R_3$, as shown below.
\begin{equation}
{L_4} = {F_3}\left( {{L_1} \oplus K_2^{'}} \right) \oplus C \oplus {K_1} \oplus {K_3}
\end{equation}
Similarly, from the decryption direction, we can get the calculation result of $L_4$ as follows.
\begin{equation}
{L_4} = {L_7} \oplus {F_5}\left( {{R_7} \oplus {F_6}\left( {{L_7}} \right) \oplus {K_6}} \right)
\end{equation}
At this point, from the encryption direction, we need to guess the subkeys $K_2^{'}$, $K_1$ and $K_3$, and in the decryption direction, we only need to guess $K_6$. By using the difference function reduction technique, we can get  the matching value $\Delta {L_4}$. Then, we just have to guess the value of the subkey $K_2^{'}$ from the encryption direction. $\Delta {L_4}$ can be calculated from the encryption direction,
\begin{equation}
\Delta {L_4} = {F_3}\left( {{L_1} \oplus K_2^{'}} \right) \oplus {F_3}\left( {L_1^{'} \oplus K_2^{'}} \right).
\end{equation}
Different matching values require different subkeys to be guessed. For the 6-round Feistel-2* construction, the subkeys that need to be guessed in the encryption and decryption directions are given in Tab. \ref{tab2}.
\begin{table}[h]
\begin{center}
\caption{6-round Feistel-2* subkeys with different matching values in encryption and decryption directions}\label{tab2}%
\begin{tabular}{@{}lllll@{}}
\toprule
 & $\Delta {L_1}$ & $\Delta {L_2}$ & $\Delta {L_3}$ & $\Delta {L_4}$ \\
\midrule
encryption    & -  & - & - & $K_2^{'}$     \\
decryption   & $K_3$, $K_4$, $K_5$, $K_6$   & $K_4$, $K_5$, $K_6$   & $K_5$, $K_6$   & $K_6$       \\
\toprule
\end{tabular}
\end{center}
\end{table}

For the 6-round Feistel-2* construction, from the Tab. \ref{tab2}, we can see that when calculating the matching value $\Delta {L_4}$ from the encryption and the decryption direction, the subkeys to be guessed are $K_2^{'}$ and $K_6$ respectively. So two differential pairs are needed to determine $K_2^{'}$ and $K_6$. We collect three pairs of plaintext and ciphertext, i.e. $\left( {L_1^{\left( i \right)}\lvert R_1^{\left( i \right)},L_7^{\left( i \right)}\lvert R_7^{\left( i \right)}} \right)$, where ${F_1}\left( {L_1^{\left( i \right)}} \right) \oplus R_1^{\left( i \right)} = C$ $\left( {1 \le i \le 3} \right)$. Supposing $L_{j + 1}^{\left( i \right)}\lvert R_{j + 1}^{\left( i \right)}$ is the $j$-th round output of ${L_1^{\left( i \right)}\lvert R_1^{\left( i \right)}}$, so we can calculate the difference function of the matching value $L_4^{\left( 1 \right)} \oplus L_4^{\left( 2 \right)}$ and $L_4^{\left( 1 \right)} \oplus L_4^{\left( 3 \right)}$, and then $K_2^{'}$ and $K_6$ can be guessed in parallel from the direction of encryption and decryption, respectively.

According to the above analysis, three pairs of plaintexts and ciphertexts $\left( {L_1^{\left( i \right)}\lvert R_1^{\left( i \right)},L_7^{\left( i \right)}\lvert R_7^{\left( i \right)}} \right)$, $\left( {1 \le i \le 3} \right)$  are collected, and  the functions are  designed as follows.
\begin{equation}
{f_1}(x_1) = {F_3}\left( {{L_1}^{(1)} \oplus x_1} \right) \oplus {F_3}\left( {{L_1}^{(2)} \oplus x_1} \right)
\end{equation}
\begin{equation}
{f_2}(x_1) = {F_3}\left( {{L_1}^{(1)} \oplus x_1} \right) \oplus {F_3}\left( {{L_1}^{(3)} \oplus x_1} \right)
\end{equation}
\begin{equation}
\begin{aligned}
{g_1}\left( x_2 \right) =& R_7^{\left( 1 \right)} \oplus R_7^{\left( 2 \right)} \oplus {F_5}\left( {L_7^{\left( 1 \right)} \oplus {F_6}\left( {R_7^{\left( 1 \right)}} \right) \oplus x_2} \right) \\ 
&\oplus {F_5}\left( {L_7^{\left( 2 \right)} \oplus {F_6}\left( {R_7^{\left( 2 \right)}} \right) \oplus x_2} \right)
\end{aligned}
\end{equation}
\begin{equation}
\begin{aligned}
{g_2}\left( x_2 \right) =& R_7^{\left( 1 \right)} \oplus R_7^{\left( 3 \right)} \oplus {F_5}\left( {L_7^{\left( 1 \right)} \oplus {F_6}\left( {R_7^{\left( 1 \right)}} \right) \oplus x_2} \right)\\ 
&\oplus {F_5}\left( {L_7^{\left( 3 \right)} \oplus {F_6}\left( {R_7^{\left( 3 \right)}} \right) \oplus x_2} \right)
\end{aligned}
\end{equation}
Then, from the encryption direction, we can get ${L_4}^{(1)}$ and ${L_4}^{(2)}$ as follows.
\begin{equation}
\begin{aligned}
\Delta {L_4}^{(1)} = {F_3}\left( {{L_1}^{(1)} \oplus K_2^{'}} \right) \oplus {F_3}\left( {{L_1}^{(2)} \oplus K_2^{'}} \right) = {f_1}(K_2^{'})
\end{aligned}
\end{equation}
\begin{equation}
\begin{aligned}
\Delta {L_4}^{(2)} = {F_3}\left( {{L_1}^{(1)} \oplus K_2^{'}} \right) \oplus {F_3}\left( {{L_1}^{(3)} \oplus K_2^{'}} \right) = {f_2}(K_2^{'})
\end{aligned}
\end{equation}
From the decryption direction, ${L_4}^{(1)}$ and ${L_4}^{(2)}$ can be got as follows.
\begin{equation}
\begin{aligned}
\Delta L_4^{\left( 1 \right)} =& R_7^{\left( 1 \right)} \oplus R_7^{\left( 2 \right)} \oplus  {F_5}\left( {L_7^{\left( 1 \right)} \oplus {F_6}\left( {R_7^{\left( 1 \right)}} \right) \oplus {K_6}} \right) \\
&\oplus {F_5}\left( {L_7^{\left( 2 \right)} \oplus {F_6}\left( {R_7^{\left( 2 \right)}} \right) \oplus {K_6}} \right) = {g_1}\left( {{K_6}} \right)
\end{aligned}
\end{equation}
\begin{equation}
\begin{aligned}
\Delta L_4^{\left( 2 \right)} =& R_7^{\left( 1 \right)} \oplus R_7^{\left( 3 \right)} \oplus {F_5}\left( {L_7^{\left( 1 \right)} \oplus {F_6}\left( {R_7^{\left( 1 \right)}} \right) \oplus {K_6}} \right)\\
&\oplus {F_5}\left( {L_7^{\left( 3 \right)} \oplus {F_6}\left( {R_7^{\left( 3 \right)}} \right) \oplus {K_6}} \right) = {g_2}\left( {{K_6}} \right)
\end{aligned}
\end{equation}

It is necessary to find $\left( x_1,x_2 \right)$ satisfying both conditions ${f_1}\left( x_1 \right){\rm{ = }}{g_1}\left( x_2 \right)$ and ${f_2}\left( x_1 \right){\rm{ = }}{g_2}\left(x_2 \right)$, where $x$ and $y$ are solutions of subkeys $K_2^{'}$ and $K_6$, respectively. Now we can use Algorithm \ref{algo3} to find $K_2^{'}$ and $K_6$. According to Theorem \ref{thm2}, the time complexity is $O({2^{n/3}})$ with $O({n2^{n/3}})$ qubits.


After finding the corresponding subkeys ${K_2^{'}}$ and ${K_6}$, by observing Table 1, we know that when the matching value is $\Delta {L_3}$, the subkeys that need to be guessed from the decryption direction are ${K_5}$ and ${K_6}$, and from the encryption direction, we can directly calculate the result. Among them, we know ${K_6}$, so we can use amplitude amplification to find ${K_5}$. By analogy, we find that when the matching values are $\Delta {L_2}$ and $\Delta {L_1}$, the constant result can also be calculated directly from the encryption direction. Therefore, we consider using the amplitude amplification algorithm when recovering the subkeys $\left\{ {{K_3},{K_4},{K_5}} \right\}$. We first consider using the amplitude amplification algorithm to find the subkey $\left\{ {{K_3},{K_4},{K_5}} \right\}$ in series. At this time, the time complexity of the search is ${2^{3n/4}}$, which is still very high. To further reduce the search  time and obtain a more efficient attack, we consider searching for subkeys separately. In this way, the time complexity of the search will be reduced to ${2^{n/4}}$.

Take the recovery of ${K_5}$ as an example, and the subkey ${K_5}$ needs to satisfy
\begin{equation}
\begin{array}{l}
{R_7} \oplus R_7^{'} \oplus {F_6}\left( {{L_7}} \right) \oplus {F_6}\left( {L_7^{'}} \right) \oplus {F_4}\left( {{L_7} \oplus {K_5} \oplus {F_5}\left( {{R_7} \oplus {K_6} \oplus {F_6}\left( {{L_7}} \right)} \right)} \right)\\
 \oplus {F_4}\left( {L_7^{'} \oplus {K_5} \oplus {F_5}\left( {R_7^{'} \oplus {K_6} \oplus {F_6}\left( {L_7^{'}} \right)} \right)} \right) = {L_1} \oplus L_1^{'}
\end{array}
\end{equation}
The left side of the equation can be represented by $f\left( {{K_5}} \right)$, and the right side of the equation can be represented by the constant $C$, then the equation is $f\left( {{K_5}} \right) = C$. Then, we use the Grover's algorithm to find the rest subkeys. Combining the above analysis, we define a function $B:{\left\{ {0,1} \right\}^{n/2}} \to \left\{ {0,1} \right\}$, for any input $x$,
\begin{equation}
B(x)  = \left\{ \begin{array}{l}
1,\begin{array}{*{20}{c}}
{}&{}
\end{array}if\begin{array}{*{20}{c}}
{}&{}
\end{array}f(x) = C\\
0,\begin{array}{*{20}{c}}
{}&{}
\end{array}otherwise
\end{array} \right.
\end{equation}

Since ${K_2^{'}}$ and ${K_6}$ have been determined before, in general, there is only one possible $x$ in ${2^{n/2}}$ to make $B(x) = 1$, i.e., the correct ${K_5}$, so the proportion of the solution is ${2^{ - n/2}}$. For function $B$, define the unitary operator ${O_B}$ as 
\begin{equation}
{O_B}\lvert  \varphi \rangle  = \left\{ \begin{array}{l}
 - \lvert  \varphi \rangle ,\begin{array}{*{20}{c}}
{}&{}
\end{array}if\begin{array}{*{20}{c}}
{}&{}
\end{array}B\left( \varphi  \right) = 1\\
\lvert  \varphi \rangle ,\begin{array}{*{20}{c}}
{}&{}&{}&{}
\end{array}if\begin{array}{*{20}{c}}
{}&{}
\end{array}B\left( \varphi  \right) = 0
\end{array} \right.
\end{equation},
and then apply Grover‘s iteration $R \approx \frac{\pi }{4}\sqrt {{2^n}} $ times. Finally ${{K_5}}$ can be recovered.

The recovery process is the same for ${{K_3}}$ and ${{K_4}}$. At this point, we can guess the subkeys $\left\{ {K_2^{'},{K_3},{K_4},{K_5},{K_6}} \right\}$. According to the definition $K_1^{'} = {F_2}\left( {{K_1} \oplus C} \right)$, $K_2^{'} = K_1^{'} \oplus {K_2}$, if we know ${K_1}$ and $K_2^{'}$, then we can get ${{K_2}}$. Now that $K_2^{'}$ is known, the remaining subkey ${{K_1}}$ can also be searched by amplitude amplification. At this time, we have obtained all the subkeys of Feistel-2* construction according to the quantum ASR attack. So the time complexity of restoring $K_1$, $K_2$, $K_3$, $K_4$, $K_5$ and $K_6$ is $O({2^{n/4}})$. In summary, the total time complexity is $O({2^{n/3}})$, the data complexity is only $O(1)$, and the number of qubits is O($n$).

\subsection{Application: Attack on 6-round Simeck32/64}\label{subsec4.3}

\begin{figure}
    \centering
    \includegraphics[width=6cm]{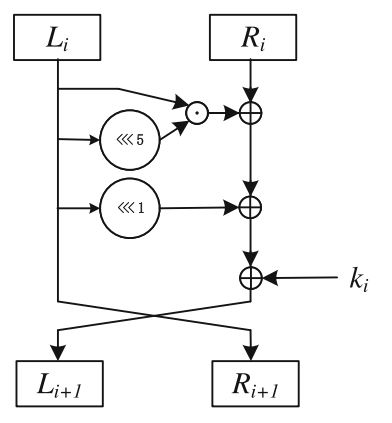}
    \caption{The round function of Simeck}
    \label{fig4}
\end{figure}

In this section, we will attack on  6-round Simeck32/64 to prove its feasibility. Simeck32/64 is an algorithm based on Feistel-2* \cite{Yang2015}. Its round function is shown in the Figure \ref{fig4}, where $x \lll c$ denotes the cyclic shift of $x$ to the left by $c$ bits, and $x \odot y$ is the bitwise AND of $x$ and $y$. Three groups of test inputs are listed in Tab. \ref{tab4}, where $f( {{L_1}^{(i)}}) \oplus {R_1}^{(i)} = C$, $C=(FFEE)_{16}$.

\begin{table}[h]
\begin{center}
\caption{Test plaintext input of 6-round Simeck32/64 (Binary)}\label{tab4}%
\begin{tabular}{@{}lll@{}}
\toprule
   & ${L_1}^{(i)}$ & ${R_1}^{(i)}$  \\
\midrule
$i=1$    & $11001101 11110101$  & $11101000 10110100$   \\
$i=2$  & $11000001 10010001$ & $1111110011011101$     \\
$i=3$   & $11010000 11000100$   & $0100111011100111$    \\
\toprule
\end{tabular}
\end{center}
\end{table}

Set the master key $K=(B0AE C7E9 C3CE E6C3)_{16}$. According to the key expansion of Simeck \cite{Kolbl2016}, then all subeys are shown in Tab. \ref{tab5}. Through Simeck encryption algorithm, we can get the corresponding ciphertext as shown in Tab. \ref{tab6}.

\begin{table}[h]
\begin{center}
\caption{Subkeys of 6-round Simeck32/64 (Binary)}\label{tab5}%
\begin{tabular}{@{}ll@{}}
\toprule
   & ${k_i}$  \\
\midrule
$i=1$    & $1011000010101110$   \\
$i=2$  & $1100011111101001$     \\
$i=3$   & $1100001111001110$     \\
$i=4$    & $1110011011000011$ \\
$i=5$  & $0000010110101001$   \\
$i=6$   & $1111111001000000$   \\
\toprule
\end{tabular}
\end{center}
\end{table}

\begin{table}[h]
\begin{center}
\caption{Test ciphertext output of 6-round Simeck32/64 (Binary)}\label{tab6}%
\begin{tabular}{@{}lll@{}}
\toprule
   & ${L_7}^{(i)}$ & ${R_7}^{(i)}$  \\
\midrule
$i=1$    & $1011 1110 0011 1010$  & $10001110 11001111$   \\
$i=2$  & $0111100111110100$ & $01010001 01110100$     \\
$i=3$   & $1011010001101000$   & $11001010 00110100$    \\
\toprule
\end{tabular}
\end{center}
\end{table}

According to our attack, we can construct ${f_1}(x_1)$, ${f_2}(x_1)$,${g_1}(x_2)$ and ${g_2}(x_2)$.

\begin{equation}
\begin{array}{l}
\begin{aligned}
{f_1}(x_1) = & ({L_1}^{(1)} \oplus {x_1} \odot ({L_1}^{(1)} \oplus {x_1} \lll 5)) \oplus ({L_1}^{(1)} \oplus {x_1} \lll 1)\\
 & \oplus ({L_1}^{(2)} \oplus {x_1} \odot ({L_1}^{(2)} \oplus {x_1} \lll 5)) \oplus ({L_1}^{(2)} \oplus {x_1} \lll 1)
\end{aligned}
\end{array}
\end{equation}

\begin{equation}
\begin{array}{l}
\begin{aligned}
{f_2}(x_1) = & ({L_1}^{(1)} \oplus {x_1} \odot ({L_1}^{(1)} \oplus {x_1} \lll 5)) \oplus ({L_1}^{(1)} \oplus {x_1} \lll 1)\\
& \oplus ({L_1}^{(3)} \oplus {x_1} \odot ({L_1}^{(3)} \oplus {x_1} \lll 5)) \oplus ({L_1}^{(3)} \oplus {x_1} \lll 1)
\end{aligned}
\end{array}
\end{equation}

\begin{equation}
\begin{array}{l}
\begin{aligned}
{g_1}({x_2}) = & {R_7}^{(1)} \oplus {R_7}^{(2)} \oplus (({L_7}^{(1)} \oplus ({R_7}^{(1)} \odot ({R_7}^{(1)} \lll 5)) \oplus ({R_7}^{(1)} \lll 1) \oplus {x_2})\\
& \odot (({L_7}^{(1)} \oplus ({R_7}^{(1)} \odot ({R_7}^{(1)} \lll 5)) \oplus ({R_7}^{(1)} \lll 1) \oplus {x_2}) \lll 5))\\
& \oplus (({L_7}^{(1)} \oplus ({R_7}^{(1)} \odot ({R_7}^{(1)} \lll 5)) \oplus ({R_7}^{(1)} \lll 1) \oplus {x_2}) \lll 1)\\
& \oplus (({L_7}^{(2)} \oplus ({R_7}^{(2)} \odot ({R_7}^{(2)} \lll 5)) \oplus ({R_7}^{(2)} \lll 1) \oplus {x_2})\\
& \odot (({L_7}^{(2)} \oplus ({R_7}^{(2)} \odot ({R_7}^{(2)} \lll 5)) \oplus ({R_7}^{(2)} \lll 1) \oplus {x_2}) \lll 5))\\
& \oplus (({L_7}^{(2)} \oplus ({R_7}^{(2)} \odot ({R_7}^{(2)} \lll 5)) \oplus ({R_7}^{(2)} \lll 1) \oplus {x_2}) \lll 1)
\end{aligned}
\end{array}
\end{equation}

\begin{equation}
\begin{array}{l}
\begin{aligned}
{g_2}({x_2}) = & {R_7}^{(1)} \oplus {R_7}^{(3)} \oplus (({L_7}^{(1)} \oplus ({R_7}^{(1)} \odot ({R_7}^{(1)} \lll 5)) \oplus ({R_7}^{(1)} \lll 1) \oplus {x_2})\\
& \odot (({L_7}^{(1)} \oplus ({R_7}^{(1)} \odot ({R_7}^{(1)} \lll 5)) \oplus ({R_7}^{(1)} \lll 1) \oplus {x_2}) \lll 5))\\
& \oplus (({L_7}^{(1)} \oplus ({R_7}^{(1)} \odot ({R_7}^{(1)} \lll 5)) \oplus ({R_7}^{(1)} \lll 1) \oplus {x_2}) \lll 1)\\
& \oplus (({L_7}^{(3)} \oplus ({R_7}^{(3)} \odot ({R_7}^{(3)} \lll 5)) \oplus ({R_7}^{(3)} \lll 1) \oplus {x_2})\\
& \odot (({L_7}^{(3)} \oplus ({R_7}^{(3)} \odot ({R_7}^{(3)} \lll 5)) \oplus ({R_7}^{(3)} \lll 1) \oplus {x_2}) \lll 5))\\
& \oplus (({L_7}^{(3)} \oplus ({R_7}^{(3)} \odot ({R_7}^{(3)} \lll 5)) \oplus ({R_7}^{(3)} \lll 1) \oplus {x_2}) \lll 1)
\end{aligned}
\end{array}
\end{equation}

We need $132+98*2^{32/3}$ qubits to encode the function, and then take the function as the input of Algorithm \ref{algo3}. Through measurement, we can get $(x_1, x_2)=(00010001011010001, 1111111001000000)$. After verification and comparison, it can be concluded that $(x_1, x_2)=({k}_2^{'}, k_6)$. Then we can use the Grover's algorithm to find the remaining keys $k_3=(1100011100110)$, $k_4=(111011011000011)$ and $k_5=(00000 10110101001)$. According to the definition, $k_1=(10110000101110)$ and $k_2=(1100011111101001)$ can also be restored.

\section{Complexity analysis}\label{sec5}

In this section, the quantum ASR attack on 6-round Feistel-2* structure is analyzed from the data complexity, query complexity, memory complexity and the number of qubits. 

As mentioned in Sect. \ref{subsec4.2}, two subkeys need to be guessed to compute matching values from the encryption and decryption directions, so we only require 3 plaintext and ciphertext pairs to attack the 6-round Feistel-2* structure. Therefore, our data complexity is $O(1)$. In addition, the query and memory complexity of Algorithm \ref{algo3} are $O({2^{n/3}})$ and $O({2^{n/3}})$ respectively, and the number of qubits is $O({n2^{n/3}})$. The query complexity of restoring $K_1-K_5$ is $O({2^{n/4}})$. Therefore, the total query complexity is $O({2^{n/3}})$, the data complexity is only $O(1)$, the memory complexity is $O({2^{n/3}})$, and the number of qubits is $O({n2^{n/3}})$. 

In the Yang et al.'s attack \cite{Yang2017}, the complexity of claw finding is not calculated, so we recalculated it. Assuming that the fastest classical claw finding algorithm (i.e., first sort the two $N=2^n$-size tables and then match them sequentially, with a query complexity of $O(NlogN)$ and memory complexity of $O(N)$) is used in the classical attack, then the complexity of the finding is $O(n{2^n})$. So the classical query complexity is $O(n{2^{n/2}})$, and the memory complexity is $O({2^{n/2}})$. It is obvious that our attack has a great improvement in query and memory complexity than the classical all-subkeys-recovery attack.

Under the Q2 model, Ito et al. \cite{Ito2019} proposed the first 4-round quantum distinguisher on Feistel ciphers in the quantum chosen-ciphertext setting and extended the distinguisher to Feistel-2* structures for key recovery attacks. The query complexity of the Ito et al.'s attack \cite{Ito2019} is $O({2^{n/4}})$ and the data complexity is $O({2^{n}})$.
Although our query complexity is a little bit higher than the chosen plaintext attack under the Q2 model, our data complexity is only $O(1)$, which is much lower than the attack by Ito et al. And our attack belongs to the Q1 model rather than the Q2 model, which means that our attack can be applied if the adversary is not allowed to access the quantum cryptographic oracle, which is more realistic to apply.

A comparison of classical and quantum attack on 6-round Feistel-2* structure be provided in Tab. \ref{tab3}.

\begin{table}[h]
\begin{center}
\caption{Comparison with previous attacks on Feistel-2* structure}\label{tab3}%
\begin{tabular}{@{}lllllll@{}}
\toprule
 & Mode & Round & Query & Memory & Data & Bits/Qubits\\
\midrule
Yang et al.'s attack\cite{Yang2017}    & Classic  & 6 & $O(n{2^{n/2}})$ & $O({2^{n/2}})$ & $O(1)$ & $O({n2^{n/2}})$    \\
Ito et al.'s attack\cite{Ito2019}    & Q2  & 6 & $O({2^{n/4}})$ & $O({2^{n/2}})$ & $O({2^{n}})$ & $O({n2^{n}})$    \\
Our attack   & Q1   & 6   & $O({2^{n/3}})$   & $O({2^{n/3}})$   & $O(1)$ & $O({n2^{n/3}})$   \\
\toprule
\end{tabular}
\end{center}
\end{table}

\section{Conclusion}\label{sec6}

As we have seen, the main advantage of our attack is the great reduction of data complexity, but the reduction of query complexity and memory complexity still needs to be improved. In this paper, a quantum ASR attack based on multi-equations quantum claw-finding is proposed, which is more realistic and has a low data complexity. Furthermore, our attack is not only applicable to 6-round Feistel-2* structure, but can be extended to r-round Feistel-2* structure.

Our work only considers fewer restraints on the round function, and whether the method is suitable for other block cipher remains to be studied. Recently, several scholars have designed the quantum circuits of AES-128 \cite{Brandon2020,Wang2022a}, and even adopted the variational quantum attack algorithm to perform the cryptanalysis of AES or AES-like structures \cite{Wang2022b}, which are all areas worthy of further study.

\bmhead{Acknowledgments}

This work is supported by the National Natural Science Foundation of China (62071240, 61802175), the Natural Science Foundation of Jiangsu Province (BK20171458).






\end{document}